\documentclass[twocolumn,letter]{jpsj2}

\def\runtitle{Neutrino Astronomies at the Horizon}
\def\runauthor{Daniele {\sc Fargion} }

\title
{The Rise of  High Energy Neutrino Astronomy at Horizon}

\author
{ Daniele {\sc Fargion}\footnote{Present address: Department of
Physics, Rome University "La Sapienza", Ple. A. Moro 2,
00185,Rome,Italy} $^{1,}$\footnote{Second footnote test.} }

\inst { Department of Physics, Rome University "La Sapienza",
Ple. A. Moro 2,
00185,Rome\\
$^1$INFN,Rome 1, }

\recdate { \today }

\abst { High Energy Neutrino Astronomy   energy windows may be
ruled by a new Upward and Horizontal Tau Air-Showers (UpTaus,
HorTaus) detectors (scintillators and optical arrays) located a
few kilometers beyond  facing High Mountains Chains, or located
above their top or  flying on Planes, on Balloons and on
Satellites facing downward to the Earth Crust Horizon. While
looking downward to the Earth seeking for Upward and Horizontal
Tau Air-Showers the same optical detectors may often capture
cosmic rays Cherenkov lights reflected or diffused by downward
Shower cosmic rays
 hitting the Earth soil (a sea,a lake, ice lands, desert or
ground). These new detectors may be built up as a circular hybrid
crown array at high quota facing to horizon  in correlation with
present largest telescopes, like Magic or Shalon or ASHRA  ones,
if able to look below the Earth edge. The UpTaus and HorTau
Astronomy  at $10^{15}$ eV up to $10^{19}$ eV energy windows may
test the largest primary neutrino fluxes in Z-Burst neutrino
model needed to solve the GZK puzzle or even the smaller but
inevitable GZK neutrino flux secondary of the same GZK cut-off.
These HorTus signals might well be observed in future EUSO
experiment.}

\kword { Neutrino, Tau, Muon, Air-Shower, Albedo, Horizon,
Detector }

\begin{document}

\sloppy \maketitle


\section{High Energy Neutrino Astronomy }

Ultra High Cosmic Rays, UHECR, diffused in isotropic arrivals at
cosmic-like features and their few cluster events  correlated to
far BL Lacs sources, seem to be originated at cosmological
distances. Moreover these UHECR events are not clustered at all to
any nearby but rarer AGN, QSRs or Known GRBs within the allowable
narrow ($10-30$ Mpc radius) volume defined by the cosmic 2.75
$K^{o}$ proton drag viscosity (the so called GZK cut-off
\cite{Greisen:1966jv}~\cite{Zatsepin:1966jv}). The recent
doublets and triplets clustering found by AGASA seem to favor
compact object (as AGN) over more exotic topological relic
models. Therefore the missing source AGN within a narrow GZK
volume is wondering. A possible remarkable correlation recently
shows that most of the UHECR event cluster point toward far away
BL Lac sources ~\cite{GorbunovTinyakovTkachevTroitsky}.
 This correlation favors a cosmic origination for UHECRs, well above
the near GZK volume. In this frame a relic neutrino mass
~\cite{Dolgov2002}~ \cite{Raffelt2002}  $m_{\nu} \simeq 0.4$ eV
 (or at least $m_{\nu} \geq 0.1 $) may solve the GZK paradox overcoming the
proton opacity: indeed a Z-Shower (or Z-burst) model may explain
that the light relic neutrino mass (larger or at least comparable
to observed splitting of neutrino masses) should diffuse into
huge hot halos whose size is a few tens of $Mpc$. This dark halo
may act as a calorimeter leading to a primary UHE neutrino (at
ZeV energies) to hit a relic one neutrino leading to a Z boson
production at resonance peak whose relativistic boosted decay
contains the nuclear secondaries
 finally observed on Earth atmosphere as UHECR
  ~\cite{FargionSalis1997}~\cite{FargionMeleSalis1999}~\cite{Weiler1999}~\cite{Yoshidaetall.1998}
~\cite{Fargionetall.2001b}~\cite{FodorKatzRingwald2002}.

The light relic neutrinos are the target (the calorimeter) where
UHE $\nu$ ejected by distant (above GZK cut-off) sources may hit
and convert (via Z boson production and decay) their energy into
nuclear UHECR.

\begin{figure}
\vspace{-2.1cm}
\centering

\includegraphics[width=6.5cm]{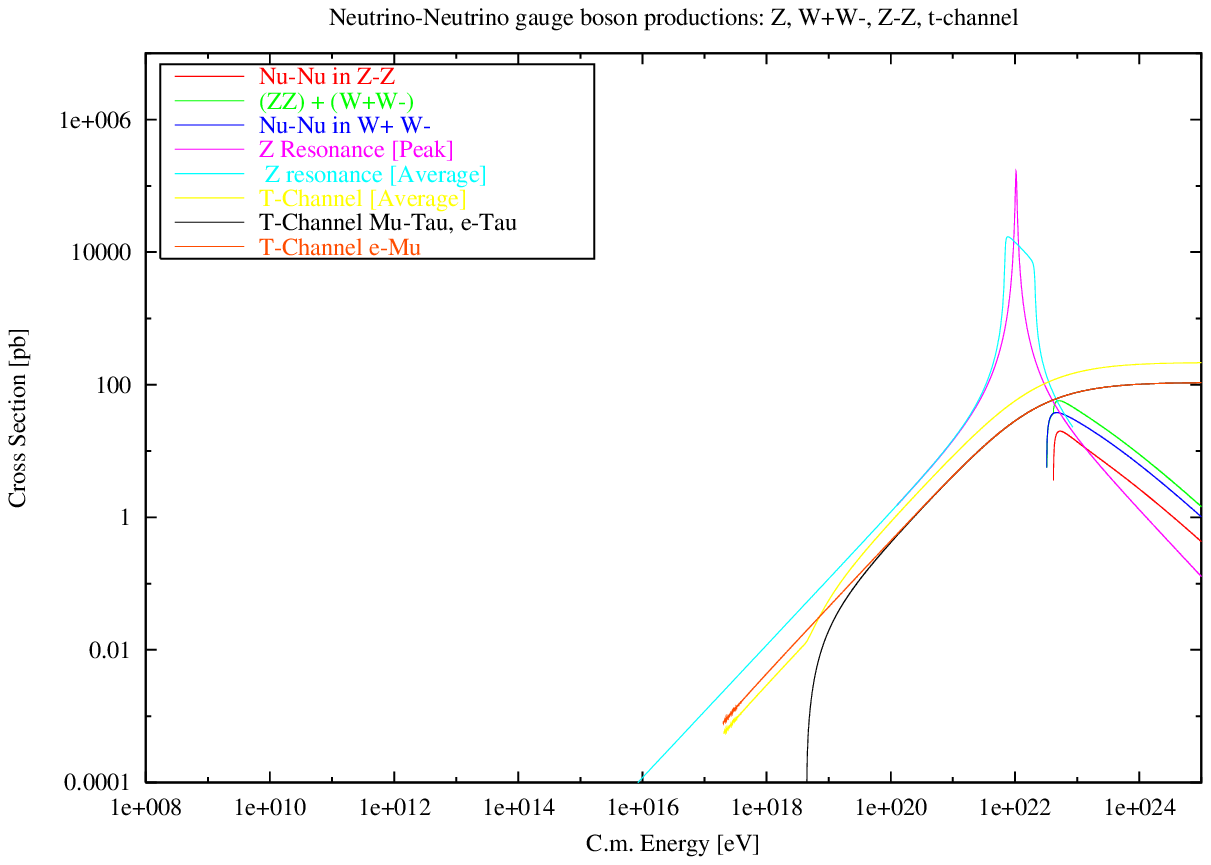}

 \vspace{-0.3cm}
\caption {The neutrino-relic neutrino cross-sections in
Laboratory (our reference) system, for a light $m_{\nu}$=$0.4$ eV
mass and a resonant incoming neutrino energy at $E_{\nu}= 10^{22}$
eV energy. The UHE neutrino and consequent Z-peak energy  will be
boosted and its decay will be smoothed in  average into the
inclined-tower peak curve; the presence of WW and ZZ channels will
guarantee a Showering also above a $2 eV$ neutrino masses. }
\label{fig:fig1}
 \vspace{0.3cm}

\centering
\includegraphics[width=6.5cm]{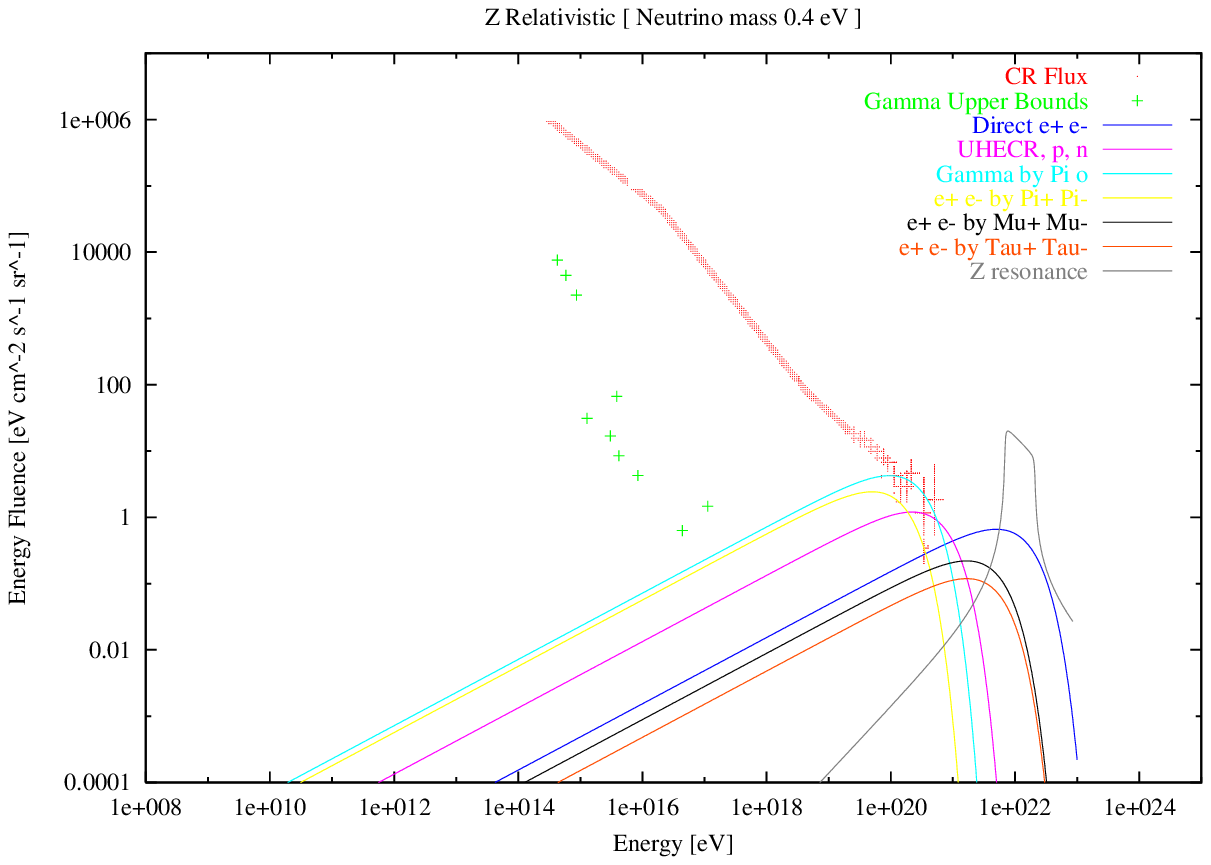}

 \vspace{-0.3cm}
\caption{Z-Showering Energy Flux distribution for different
    Z- decay channels assuming a light (fine tuned)  relic neutrino mass
    $m_{\nu} = 0.4 eV$ and a resonant incoming neutrino energy at $10^{22}$ eV energy.
     In this case    following recent solar and atmospheric neutrino mass splitting,
    all the relic neutrino masses are nearly  degenerate.
    This occurs  also  for a relic neutrino mass at $m_{\nu} = 0.2 eV$
     (in agreement with Cosmic Back ground \cite{Allen} )
     and a resonant incoming neutrino energy at $E_{\nu}$$ \simeq 2 \cdot 10^{22}$ eV.
      In figure are shown  total energy percentage
distributions  into neutrino,nucleons and their antiparticles,
neutral and charged pions  and consequent $\gamma$, $e^+$,$e^-$
particles both from hadronic and leptonic Z, $WW,ZZ$ channels. We
calculated the electro-magnetic contribution due to the t-channel
$\nu_i \nu_j$ interactions as described in previous figure. Most
of the $\gamma$ radiation will be degraded around PeV energies by
$\gamma \gamma$ pair production with cosmic 2.75 K BBR, or with
cosmic radio background polluting both MeV up to GeV energy
background. For large neutrino flux $\Phi_{\nu}$$\simeq 10^3 eV
cm^{-2}cm^{-1} sr^{-1}$ at $10^{22}$ eV these curves are
marginally consistent with present bounds $\Phi_{\nu}$ while
fitting  UHECR at AGASA fluence, for a tiny neutrino mass
$m_{\nu}\simeq 0.4 eV$. For lower HIRES data the needed flux
reduces to $\Phi_{\nu}$$\simeq 2\cdot 10^2 eV cm^{-2}cm^{-1}
sr^{-1}$ well compatible with most $\Phi_{\nu}$ bounds.}
\label{fig:fig2}
\end{figure}


\begin{figure}
\centering
\includegraphics[width=6.5cm]{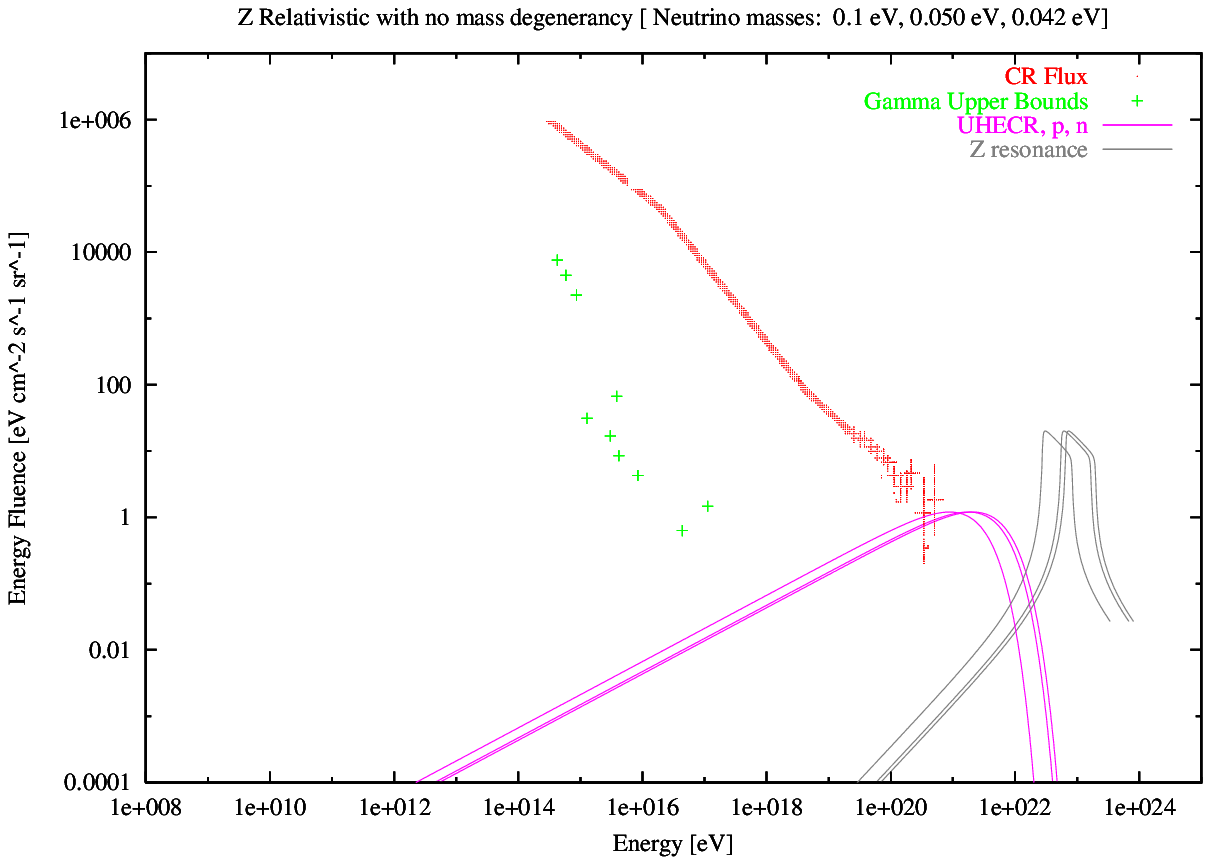}
\caption{Z-Showering Energy Flux distribution for  a non
degenerated twin light  relic neutrino masses near atmospheric
splitting mass values $m_{\nu} = 0.1 eV$, $m_{\nu}=0.05eV$,
$m_{\nu} = 0.042 eV$, with incoming $E_{\nu}$$ \simeq 4\cdot
10^{22}-8\cdot 10^{22}$ eV . Because of small solar neutrino mass
splitting there may be a near overlapping at lowest (as shown in
present figure) or at higher curve doubling one of their
intensity.} \label{fig:fig3}
\vspace{0.4cm}
\centering
\includegraphics[width=6.5cm]{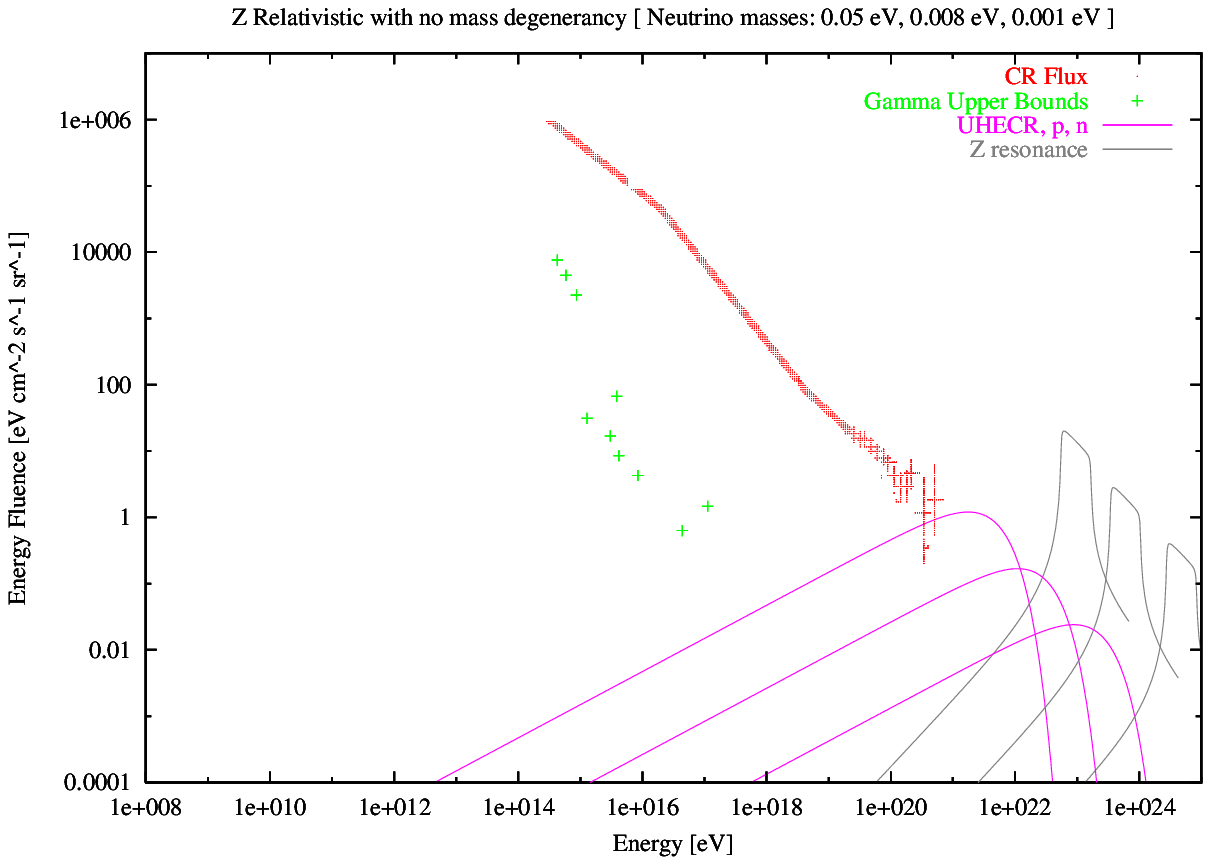}
\caption{As in Fig.\ref{fig:fig3} for not degenerated neutrino
mass $m_{\nu 1} = 0.05 eV$ (as atmospheric mass split),  $m_{\nu
2} = 0.008 eV$ (as solar mass split), $m_{\nu 3} = 0.001 eV$
(nearly massless ultra-relativistic $\nu_r$) assuming an incoming
UHE neutrino energy spectra extending up to GUT energies:
$E_{\nu}$$ \simeq 8\cdot 10^{22}-4\cdot 10^{24}$ eV .Only
hadronic shower are shown for sake of simplicity.  A suppression
for lightest neutrino densities has been assumed. }
\label{fig:fig4}
\end{figure}
\vspace{0.3cm}

These light neutrino masses do not solve the galactic or cosmic
dark matter problem but they are well consistent with atmospheric
neutrino mass splitting ($\triangle m_{\nu} \geq 0.05$ eV) and in
fine tune with more recent neutrino double beta decay experiment
mass claim $m_{\nu} \simeq 0.4$ eV
~\cite{Klapdor-Kleingrothaus:2002ke} . The neutrino mass existence
has been also motivated by old and recent solar neutrino
oscillation evidences
\cite{Gallex92}~\cite{Fukuda:1998mi}~\cite{SNO2002} and most
recent claims by KamLAND ~ \cite{Kamland2002} of anti-neutrino
disappearance  (all in agreement within a Large Mixing Angle
neutrino model and $\triangle {m_{\nu}}^2 \sim 7 \cdot
10^{-5}{eV}^2$). In this Z-WW Showering for light neutrino mass
models large fluxes of UHE $\nu$ are necessary
\cite{FargionMeleSalis1999}~\cite{Yoshidaetall.1998}~
\cite{Fargionetall.2001b}~\cite{FodorKatzRingwald2002}~\cite{Kalashev:2002kx}
or higher neutrino relic density  or better clustering are needed
~\cite{Fargionetall.2001b}~\cite{Singh-Ma} : indeed a heaviest
neutrino mass  $m_{\nu} \simeq 1.2-2.2$ eV while still being
compatible with known bounds (but marginally with more severe
WMAP indirect limits), might better gravitationally cluster
leading to denser dark local-galactic halos and lower neutrino
fluxes ~\cite{Fargionetall.2001b}~\cite{Singh-Ma} . It should
remarked that in this scenario  $m_{\nu}\geq eVs$ the main
processes leading to UHECR above GZK are mainly the WW-ZZ and the
t-channel interactions
~\cite{FargionMeleSalis1999}~\cite{Fargionetall.2001b} . It
should be noticed that a smaller UHECR flux as low as the HIRES
last claims would relax the needed UHE neutrino fluxes
$\Phi_{\nu}$  at  energies, $E_{\nu}\simeq 10^{22} eV$, at fluence
below $\Phi_{\nu}$ $\leq 10^3$ eV $cm^{-2} s^{-1}  sr^{-1}$,
making it well consistent with all known neutrino fluence bounds.
The neutrino-neutrino cross-sections at boosted incoming neutrino
energies in laboratory system are summirized in
Fig.\ref{fig:fig1}. The consequent showering at all main hadronic
and electromagnetic channels are described for a possible
degenerated neutrino mass $m_{\nu} = 0.4 eV$ Fig.\ref{fig:fig2}
(not consistent with most severe WMAP neutrino mass bounds). The
case for lighter neutrino masses comparable to atmospheric ones
are considered Fig.\ref{fig:fig3}; the case for extremely light
and non degenerated neutrino mass splitting is  also considered in
Fig.\ref{fig:fig4}; see also \cite{Fargion2002e} for description
of neutrino densities and detailed decay channel chains. Common
critic to the Z-Shower model is the
 absence in UHECR at $E_{UHECR} \simeq 10^{19}$ eV of large detectable  $\gamma$
 signal (at least at $50\%$ level of UHECR) contrary to the apparent prediction
 of dominat $\gamma$ fluxes shown in Fig.\ref{fig:fig2} ,\ref{fig:fig3}.
  However as the neutrino mass goes toward
 lighter and lighter values it requires wider and wider
 hot dark neutrino halos.  Therefore the widest relic neutrino halo may encompass
 the same GZK-cut off radius. In this case while UHECR nucleon produced by Z-Shower in GZK volumes will not
 suffer in its propagation lenght $R_{UHECR}$ of any severe cut off ($R_{UHECR} \simeq 10^3 Mpc$) its $\gamma$
 UHECR companion propagation lenght will be
exponentially suppressed ($R_{\gamma} < 6 Mpc$) by the cosmic
radio opacity leading at least to a cubic volume suppression
$$({\frac{R_{\gamma}}{R_{UHECR}})^{3} \ll 1}$$
 This   explains the suppression of any large
 $\gamma$ signal at $E_{UHECR} \simeq 10^{19}$ eV in Z-Shower model.

  Therefore UHE neutrino in Z-shower model might
be an abundant primary of UHECR. Vice-versa, even ignoring  the
GZK puzzle and rejecting the Z-Shower model, the same
extragalactic UHECR should be at least source during their
propagation (by the same photo-nuclear interaction responsible of
GZK cut-off)  of the so called Greisen GZK neutrino secondaries
around $10^{19}$ eV energies. The independent measure of these
neutrino fluxes (either primary as in Z-Burst or secondary as in
a naive GZK case) is therefore of great scientific urgency.
Because of the small neutrino flux and of the poor weak
interactions this highest energy neutrino astronomy call for
large detector volumes, much larger than SK or SNO ones for $MeV$
energy solar neutrino. Last three decades were devoted to the
study and to the development  of a cubic Km$^3$ underground
seeking to track the Cherenkov radiation emitted  by relativistic
muons (produced by their neutrinos) either in ice (AMANDA) or in
water (Baikal, Nestor,Antares).

\section{  $\tau$ Air-Showering for $\nu$ Astronomy }

\begin{figure}
\centering
\includegraphics[width=6.5cm]{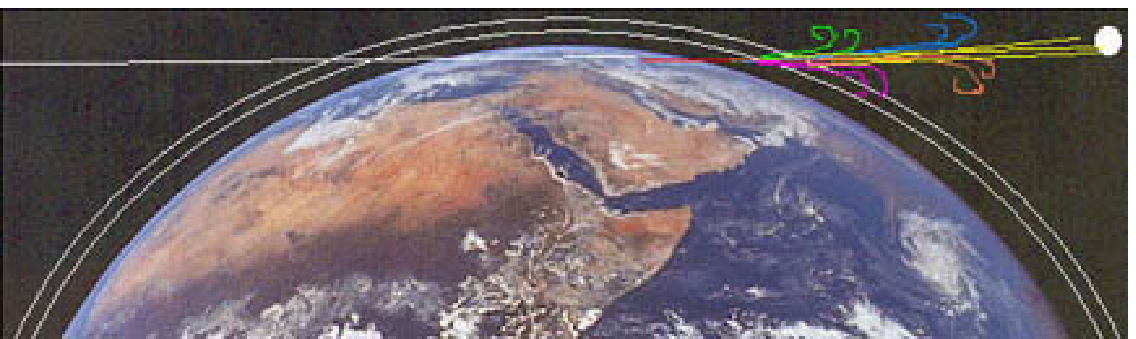}
\caption { As above  Horizontal Upward Tau Air-Shower (HORTAUS)
originated by UHE neutrino skimming the Earth: fan-like jets due
to geo-magnetic bending  shower at high quota ($\sim 23-40 km$):
they may be pointing to an orbital satellite detector . The
Shower tails may be also observable by EUSO just above it.}
\label{fig:fig6}
\vspace{0.2cm}
\centering\includegraphics[width=6.5cm]{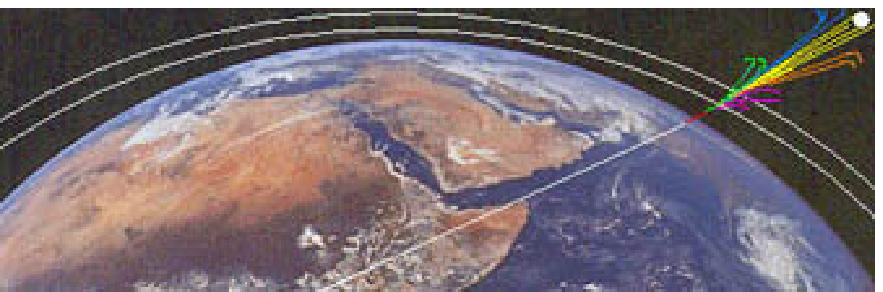}
\caption {A very schematic Upward Tau Air-Shower (UPTAUs)  and its
open fan-like jets due to geo-magnetic bending at high quota
($\sim 20-30 km$). The gamma Shower may be pointing to an orbital
detector. Its
 vertical Shower tail may be spread by
geo-magnetic field into a thin eight-shape beam observable  by
EUSO  as a small blazing oval (few dot-pixels) aligned orthogonal
to the local magnetic field .} \label{fig:fig7}
 \vspace{-0.5cm}
\end{figure}

\begin{figure}
\centering\includegraphics[width=7cm]{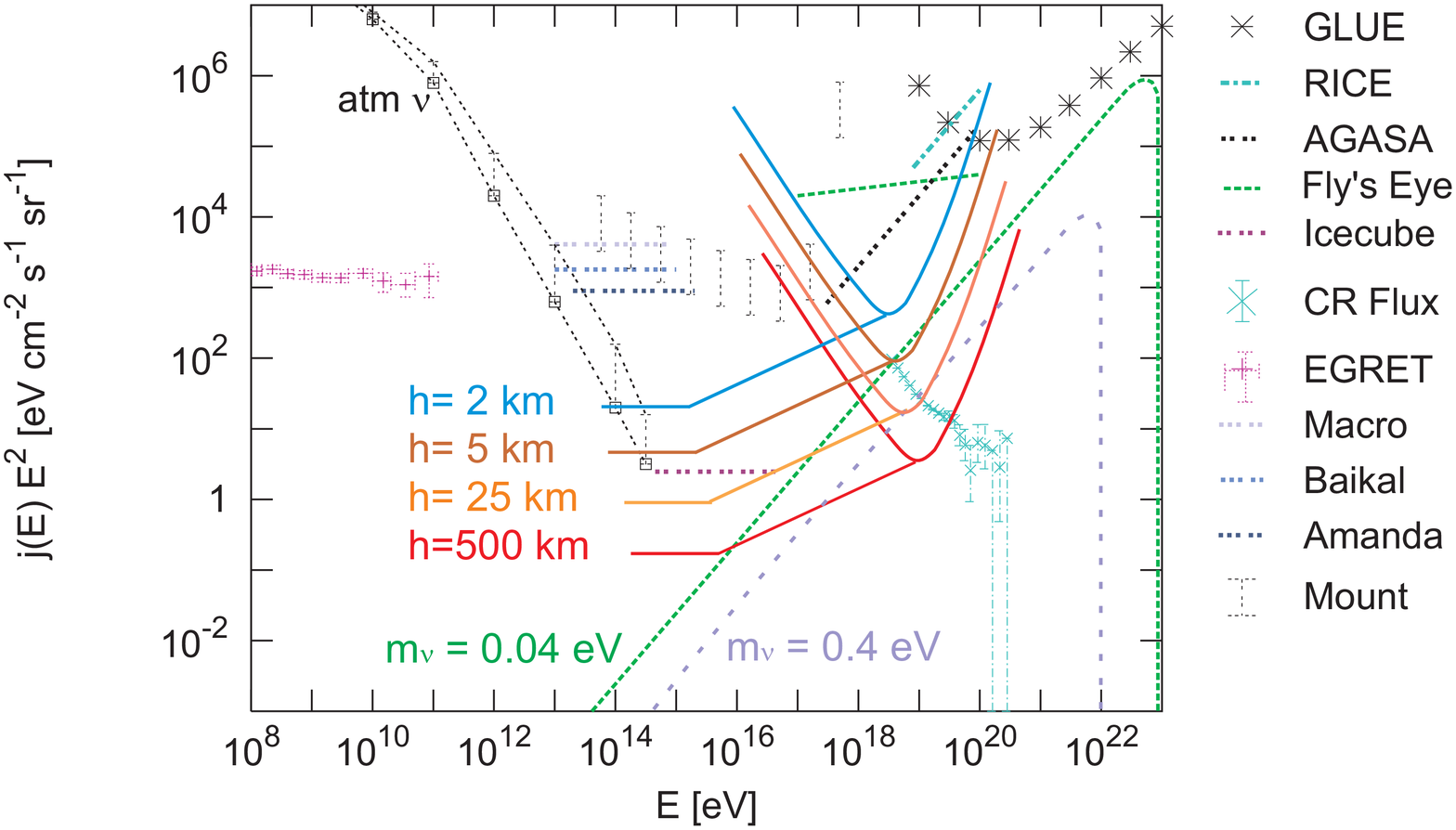}

\caption {UPTAUs (lower bound on the center) and HORTAUs (right
parabolic  curves)  sensibility at different observer heights h
($2,5,25,500 km $) looking at horizon toward Earth seeking upward
Tau Air-Showers adapted over a present neutrino flux estimate in
Z-Shower model scenario for light ($0.4-0.04$ eV) neutrino masses
$m_{\nu}$; two corresponding density contrast for relic light
neutrino masses has been assumed; the lower parabolic bound
thresholds are at different operation height, in Horizontal
(Crown) Detector facing toward most distant horizon edge; these
limits are fine tuned (as discussed in the text) in order to
receive Tau in flight and its
 Shower in the vicinity of the detector; we are assuming a
duration of data records of a decade comparable to the BATSE
record data . The parabolic bounds on the EeV energy range in the
right sides are nearly un-screened by the Earth opacity while the
corresponding UPTAUs bounds  in the center below suffer both of
Earth opacity as well as of a consequent shorter Tau interaction
lenght in Earth Crust, that has been taken into account. }
\label{fig:fig10}
\end{figure}

\begin{figure}
\centering
\includegraphics[width=8cm]{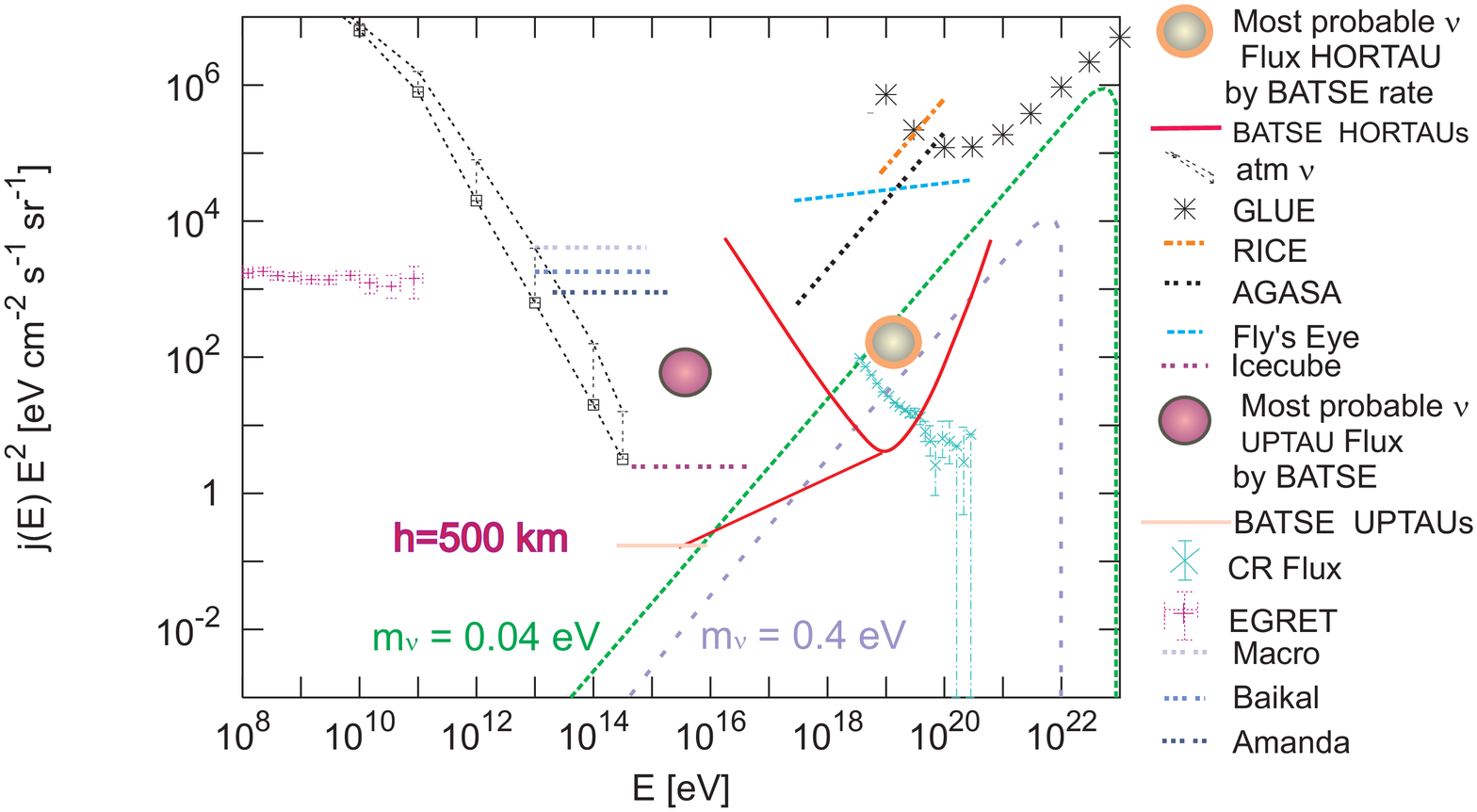}
\caption {Neutrino Flux derived by  BATSE Terrestrial Gamma
Flashes assuming them as $\gamma$ secondaries of upward Tau
Air-showers. These fluxes are estimated using the data from
Terrestrial Gamma Flash (1991-2000) normalized during  their most
active trigger and TGF hard activities. The UPTAUs and HORTAUs
rate are normalized assuming that the events at geo-center angle
above $50^o$ might be of HORTAU nature. } \label{fig:fig19}
\end{figure}


%
\begin{figure}
\centering
\includegraphics[width=7cm]{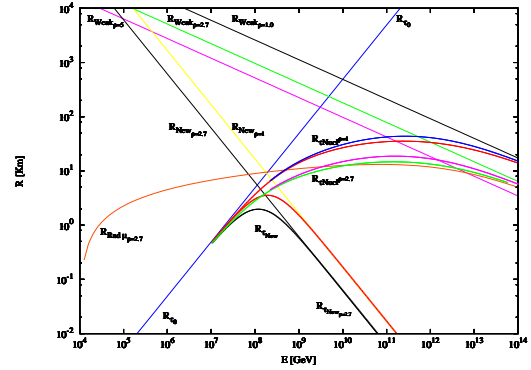}
\caption {Lepton $\tau$ (and $\mu$) Interaction Lengths for
different matter densities: $R_{\tau_{o}}$ is the free $\tau$
length,$R_{\tau_{New}}$ is the New Physics TeV Gravity interaction
range at corresponding densities,$R_{\tau_{Nucl}\cdot{\rho}}$ , is
the combined $\tau$ Ranges keeping care of all known interactions
and lifetime and mainly the photo-nuclear interaction. There are
two slightly different split curves (for each density) by two
comparable approximations in the interaction laws. Note also the
neutrino interaction lenghts above lines $R_{Weak{\rho}}=
L_{\nu}$ due to the electro-weak interactions at corresponding
densities.} \label{fig:fig8}
\end{figure}

A totally New Tau Neutrino Telescope at PeV-EeV energies has been
considered more recently
\cite{Fargionetall.1999}~\cite{Fargion2000-2002}~\cite{HouHuang2002}
based on the use of a mountain as the target and the escaping Tau
secondary and its decay in flight  as the main amplifier of the
UHE neutrino track. UHE $\nu_{\tau}$ are abundantly produced by
flavour oscillation and mixing from muon (or electron) neutrinos,
because of the large galactic and cosmic distances respect to the
neutrino oscillation ones (for already known neutrino mass
splitting)~\cite{Fargion2000-2002} . Therefore the UHE Neutrino
may interact on Mountain Chains or Earth Crust leading to UHE Tau
whose decay in flight and consequent air-shower may be observed
along or near the wide shower cone. This Astronomy is based on
$\nu_{\tau}$,$\overline{\nu_{\tau}}$ interactions  at PeV-GZKs
($E_{\nu_{\tau}}\simeq $$10^{15}-10^{19}$ eV) energies as well as
on $\overline{\nu_{e}}$ interactions (in resonance with $W^-$
gauge boson mass at $6.3$ PeV), hitting common electron in the
mountain rock. The  $W^-$ decay branching ratio contains within
$10\%$ The UHE $\overline{\nu_{\tau}}$ and $\tau$ birth whose
propagation and escape from the mountain as well as  its decay in
flight may come as the source of an amazing Horizontal Showering
from the Ande mountain Chain toward AUGER
\cite{Fargionetall.1999}~\cite{Fargion2000-2002}~\cite{Bertouetall.2002}
or by Horizontal (so called Earth-Skimming
\cite{Fargionetall.1999}~\cite{Fargion2000-2002}~\cite{Fengetall.2002})
showers flashing upward to new Mountain Observatory or present
and future Gamma Detectors in high altitude (mountain, Balloon,
Air-planes) or orbit Space
\cite{Fargion2001a}~\cite{Fargion2001b}~\cite{Fargion2000-2002} .
In this frame Tau upward Showers (UPTAUs) may have already blaze
the most sensible and celebrated GRO Observatory, on Space for
nearly a decade, leading to a rare amazing $\gamma$ showers named
as Up-going Terrestrial Gamma Flashes,TGF,  by BATSE experiment
\cite{Fargion2000-2002}~\cite{Fargion2002e} . Also Horizontal Tau
Showers (HORTAUs), at higher energies may hit the satellite from
the Horizon. Indeed it has been noticed possibly early signals in
correlation with EeV galactic anisotropy discovered by AGASA and
the EGRET galactic plane. Their deduced fluxes
~\cite{Fargion2002e}  are just at the edge of present AMANDA II
thresholds. Therefore AMANDA II may already confirm (within a few
year) these UPTAUs and HORTAUs fluxes and their  sources.

\section{EUSO experiment for UHE GZK $\nu$}

There is more and more expectation also on new space experiment
like EUSO outcoming project. This experiment, while monitoring at
dark, downward to the Earth, a wide atmosphere layers, may
discover, among common downward Ultra High Energy Cosmic Rays,
UHECR showers, also first High Energy Neutrino-Induced Showers.
These events are either originated in Air (EUSO Field of View) or
within a widest Earth Crust ring crown leading to ultra high
energy Tau whose decay in flight produce up-ward and horizontal
showers. Upward PeVs neutrinos, born on air and more probably
within a  thin Earth Crust layer, may shower rarely and blaze to
the EUSO detectors. Other higher energy neutrino may cross
horizontally hitting either the Earth Atmosphere or the Earth
Crust:  most of those vertical downward neutrinos, interacting on
air, should be  drown in the dominant noise of downward UHECR
showers. The effective target Masses originating HORTAUs seen by
EUSO may reach (on sea), for most realistic regime ( at energy
$E_{\nu_{\tau}} = 1.2 \cdot 10^{19}$ eV) a huge ring volume
$\simeq 2360$ $km^3$. The consequent HORTAUS event rate (even at
$10\%$ EUSO duty cycle lifetime) may well test the expected
Z-Burst models by at least a hundred of yearly events. However,
even rarest but inescapable GZK neutrinos (secondary of photopion
production of observed cosmic UHECR) should be discovered in a
dozen of horizontal upward shower events; in this view an
extension of EUSO detectability up to $\sim E_{\nu}\geq
10^{19}$eV threshold is mandatory. A wider collecting EUSO
telescope (3m diameter) should be considered.
 This new Neutrino
$\tau$ detector will be (at least) complementary to present and
future, lower energy, $\nu$ underground  $km^3$ telescope
projects (from AMANDA, Baikal, ANTARES, NESTOR, NEMO, IceCube). In
particular Horizontal Tau Air shower may be naturally originated
by UHE $\nu_{\tau}$ at GZK energies crossing the thin Earth Crust
at the Horizon showering far and high in the atmosphere
~\cite{Fargion2000-2002} ~\cite{Fargion2001a}
~\cite{Fargion2001b} ~\cite{Bertouetall.2002}~
\cite{Fengetall.2002} .



Let us first consider the last kind of Upward $\tau$ signals due
to their interaction in Air or in Earth Crust (UPTAUs). The Earth
opacity will filter mainly  $10^{14}\div{10^{16}}$eV upward events
~\cite{Gandhietall.1998} ~\cite{Halzen1998}
~\cite{BecattiniBottai2001} ~\cite{Duttaetall.2001}~
\cite{Fargion2000-2002} ; therefore only the direct $\nu$ shower
in air or the UPTAUs around $3$ PeVs will be able to flash toward
EUSO in a narrow beam ($2.5 \cdot 10^{-5}$ solid angle) jet
blazing apparently at $10^{19}\div{10^{20}}$eV energy. The shower
will be opened in a fan like shape and it will emerge from the
Earth atmosphere spread as a triplet or multi-dot signal aligned
orthogonal to local terrestrial magnetic field.


\begin{figure}
 \centering
\includegraphics[width=7cm]{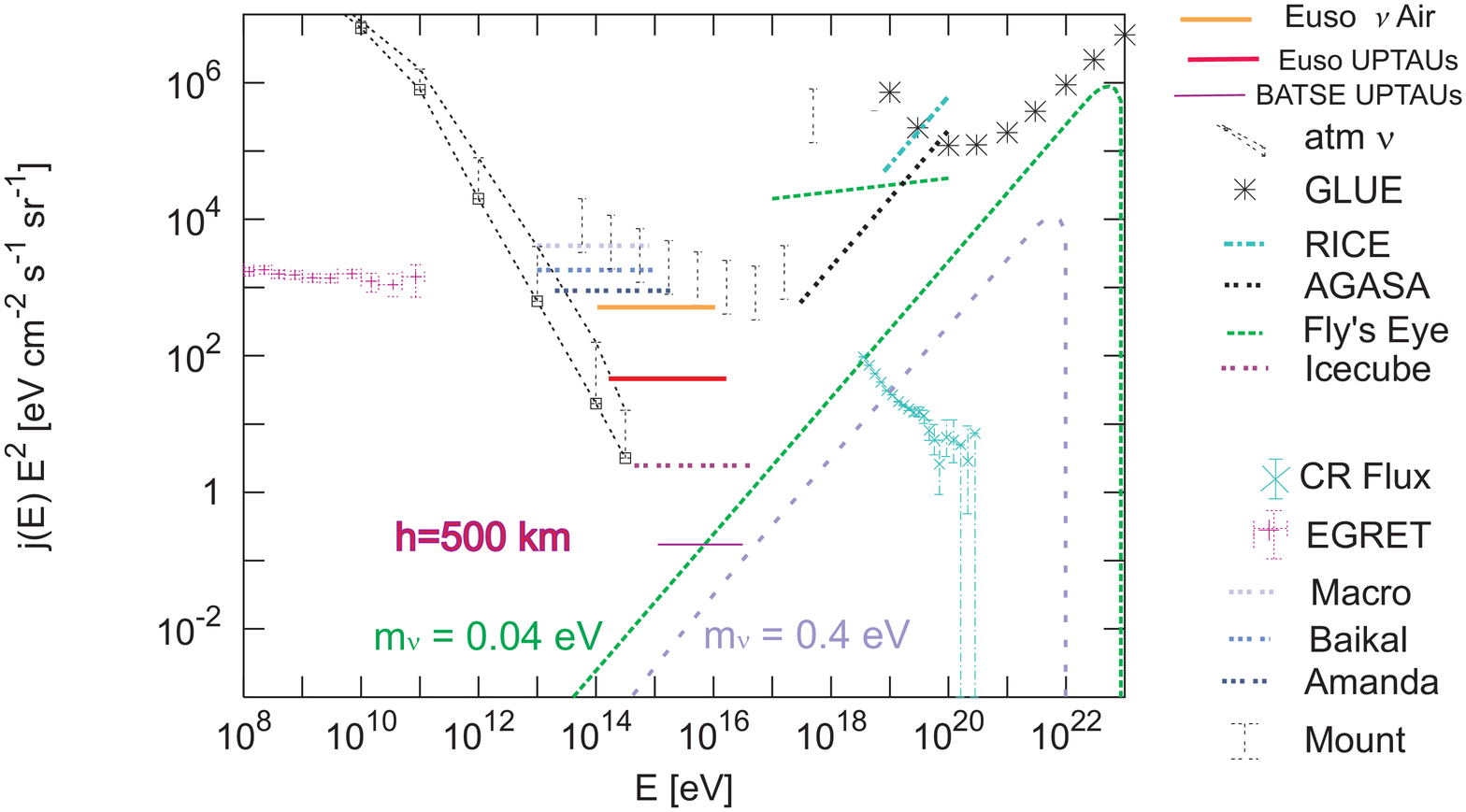}
\caption {Upward Neutrino Air-Shower and Upward Tau Air-shower,
UPTAUs, Gamma and Cosmic Rays Fluence Thresholds and bounds in
different energy windows for different past and future detectors.
The UPTAUs threshold for EUSO has been estimated for a three year
experiment lifetime. BATSE recording limit is also shown from
height $h = 500km$ and for ten year record. Competitive experiment
are also shown as well as the Z-Shower expected spectra in light
neutrino mass values $m_{\nu} = 0.4, 0.04$ eV. } \label{fig:fig9}
\end{figure}



This rare multi-dot $polarization$ of the outcoming shower will
define a characteristic  signature easily to be revealed. However
the effective observed air mass by EUSO is not $\ 10\%$ (because
duty cycle) of the inspected air volume $\sim 150 km^3$, but
because of the narrow blazing shower cone it corresponds to only
to $3.72\cdot 10^{-3}$ $km^3$. The target volume  increases for
upward neutrino Tau interacting vertically in Earth Crust in last
matter layer (either rock or water), making upward relativistic
$\simeq 3 PeVs$ $\tau$ whose decay in air born finally an UPTAUs;
in this case the effective target mass is (for water or rock)
respectively $5.5\cdot 10^{-2}$$km^3$ or $1.5 \cdot10^{-1}$
$km^3$.



The characteristic neutrino interaction are partially summirized
in Fig.\ref{fig:fig8}. The consequent $\tau$ and $\mu$
interactions lenght are also displayed. These volume are not
extreme. The consequent foreseen thresholds for UPTAUs are
summirized for $3$ EUSO years of data recording in
Fig.\ref{fig:fig9}. The UPTAUs signal is nearly $15$ times larger
than the Air-Induced Upward  $\nu$ Shower hitting Air. A more
detailed analysis may show an additional factor three (due to the
neutrino flavours) in favor of Air-Induced Showers, but the more
transparent role of PeV multi-generating upward $\nu_{\tau}$
while crossing the Earth, makes the result summirized in
Fig.\ref{fig:fig9}. The much wider acceptance of BATSE respect
EUSO and the consequent better threshold (in BATSE) is due to the
wider angle view of the gamma detector, the absence of any
suppression factor as in EUSO duty cycle, as well as the $10$
(for BATSE) over $3$ (for EUSO) years assumed of record
life-time. Any minimal neutrino  fluence $\Phi_{\nu_{\tau}}$ of
PeVs energetic neutrino:  $ \Phi_{\nu_{\tau}}\geq 10^2 eV cm^{-2}
s^{-1}$ might be detectable by EUSO.



\subsection{Downward and Horizontal UHECRs in EUSO}

Let us now briefly reconsider the nature of common Ultra High
Cosmic Rays (UHECR) showers. Their  rate  will offer a useful test
for any additional UHE neutrino signals. Let us assume for sake
of simplicity a characteristic opening angle of EUSO telescope of
$30^o$ and a nominal satellite  height of $400$ km, leading to an
approximate atmosphere area under inspection of EUSO $\sim 1.5
\cdot 10^5 km^2$. Let us discuss the UHECR shower: It has been
estimated (and it is easy to verify)  a $\sim 2\cdot10^{3}$
event/year rate above $3\cdot10^{19}$ eV. Among these "GZK" UHECR
(either proton, nuclei or $\gamma$) nearly $7.45\%\approx 150$
event/year will shower in Air Horizontally with no Cherenkov hit
on the ground. The critical angle $6.7^o$ corresponding to
$7.45\%$ of all the events, is derived from first interacting
quota (here assumed for Horizontal Hadronic Shower near $44$ km
following \cite{Fargion2000-2002}~
\cite{Fargion2001a}~\cite{Fargion2001b}) . Indeed the
corresponding horizontal edge critical angle $\theta_{h}$ $=$
$6.7^o$ below the horizon ($\pi{/2}$) is found (for an
interacting height h near $44$ km):

\begin{center}
$ {\theta_{h} }={\arccos {\frac {R_{\oplus}}{( R_{\oplus} +
h_1)}}}\simeq 1^o \sqrt{\frac {h_{1}}{km}} $
\end{center}


\begin{figure}
\centering
\includegraphics[width=6.5cm]{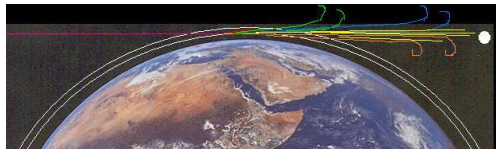}
 \caption {A  schematic Horizontal High
Altitude Shower (HIAS) suffering a geo-magnetic bending at high
quota ($\sim 44 km$). These expected events may mimic HORTAUs. The
Shower may point to a satellite. The HIAS Showers is open and
forked in five (or at least two-three main component):
($e^+,e^-,\mu^+,\mu^-, \gamma $, or just a positive-negative twin
jet ); these  multi-finger tails may be recognized by their split
tails.} \label{fig:fig5}
 \vspace{-0.2cm}
\end{figure}


\begin{figure}\centering\includegraphics[width=8cm]{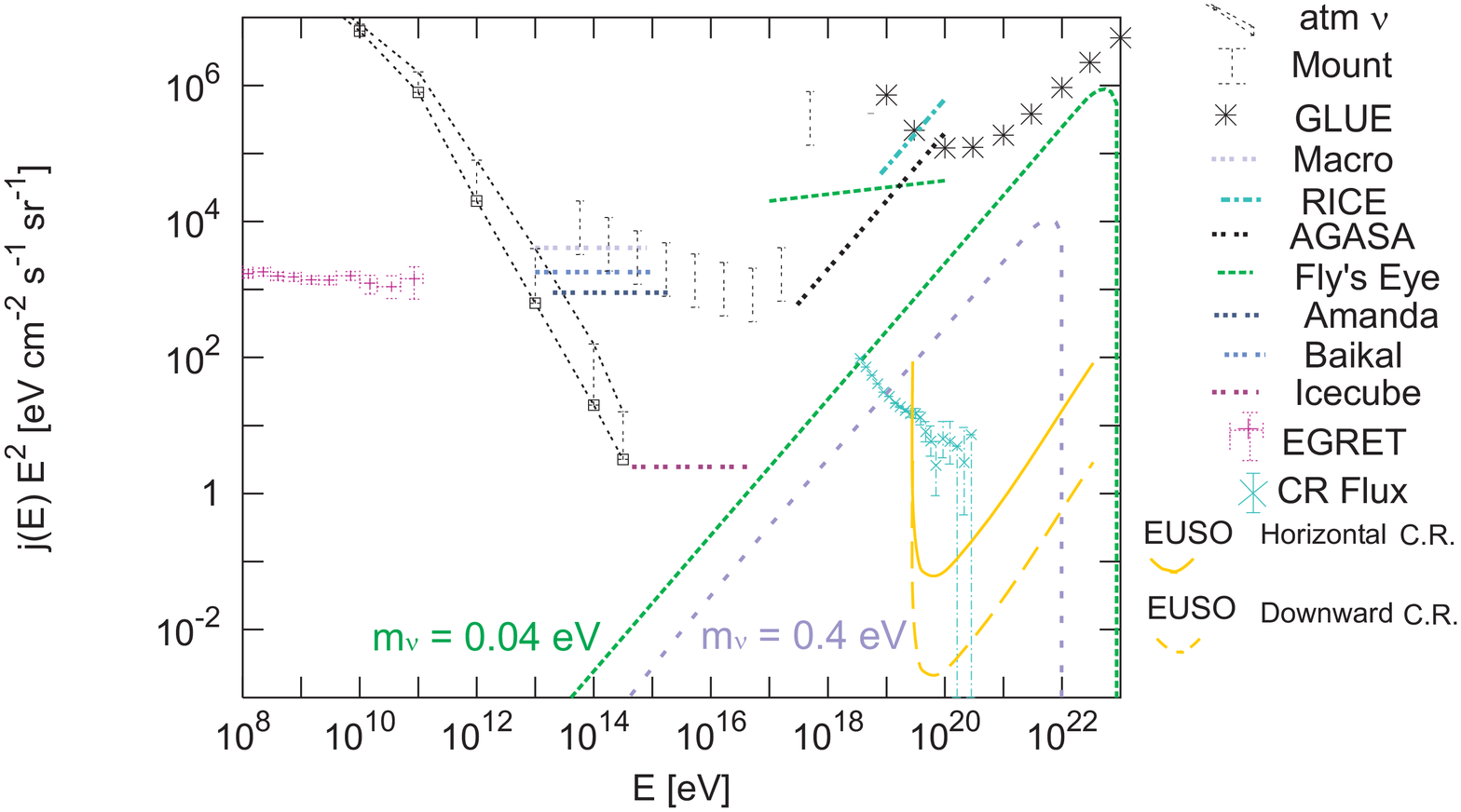}
\caption {Neutrino, Gamma and Cosmic Rays Fluence Thresholds and
bounds in different  energy windows. The Cosmic Rays Fluence
threshold for EUSO has been estimated  for a three year
experiment lifetime. The paraboloid bound shape threshold may
differ upon the EUSO optics and acceptance. Competitive
experiment are also shown as well as the Z-Shower expected
spectra in light mass values.} \label{fig:fig11}
\end{figure}



\begin{figure}\centering\includegraphics[width=8cm]{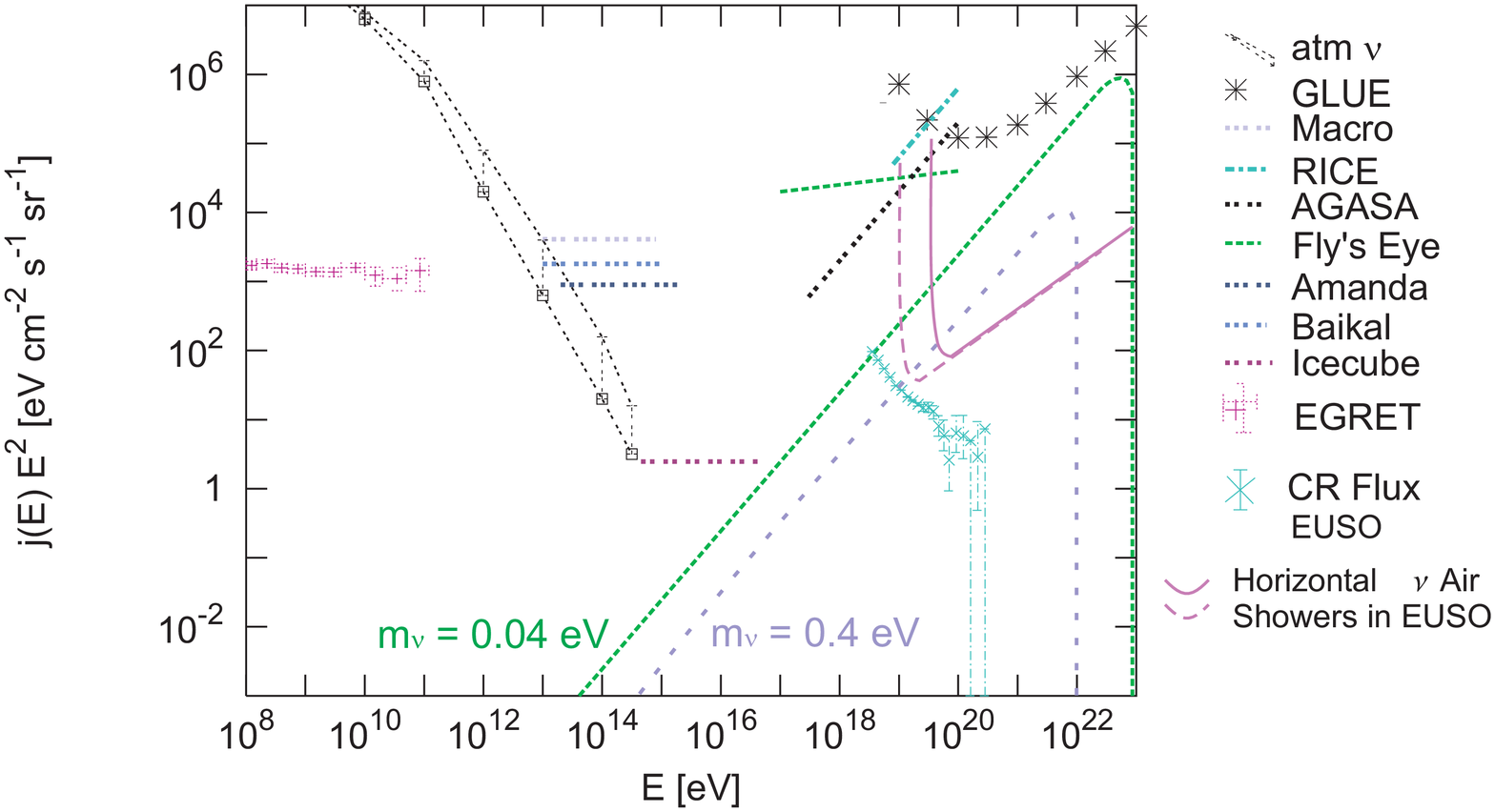}
\caption {EUSO thresholds for Horizontal and Vertical Downward
Neutrino Air induced shower over other $\gamma$, $\nu$ and Cosmic
Rays (C.R.) Fluence  and bounds. The  Fluence threshold for EUSO
has been estimated for a three year experiment lifetime.
Competitive experiments are also shown as well as the Z-Shower
expected spectra in light neutrino mass values ($m_{\nu} = 0.04,
0.4$ eV). } \label{fig:fig12}
\end{figure}

These Horizontal High Altitude Showers (HIAS)~
\cite{Fargion2001a}~ \cite{Fargion2001b} , will be able to define
a new peculiar showering, mostly very long (hundred kms) and bent
and forked (by few or several degrees) by local geo-magnetic
fields. The total UHECR above $3\cdot10^{19}$ eV will be $\sim
6000$ UHECR and $\sim 450$ Horizontal Shower within 3 years;
these latter horizontal signals are relevant because they may
mimic Horizontal  induced $\nu$ Air-Shower, but mainly at high
quota ($\geq 30-40 km$) and down-ward. On the contrary UHE
neutrino tau showering, HORTAUs, to be discussed later, are also
at high quota ($\geq 23 km$), but  upward-horizontal. Their
outcoming angle will be ($\geq 0.2^o-3^o$) upward. Therefore a
good angular ($\leq 0.2-0.1 ^o$) resolution to distinguish
between the two signal will be a key discriminator. While
Horizontal UHECR are an important piece of evidence in the UHECR
calibration and its GZK study , at the same time they are a
severe back-ground noise competitive with Horizontal-Vertical GZK
Neutrino Showers originated in Air, to be discussed below. However
Horizontal-downward UHECR are not confused with upward Horizontal
HORTAUs by UHE neutrinos to be summirized in last section. Note
that Air-Induced Horizontal UHE neutrino as well as all down-ward
Air-Induced UHE $\nu$ will shower mainly at lower altitudes
($\leq 10 km$) ; however they are respectively only a small
($\leq 2\% $, $\leq 8\%$) fraction than HORTAUs showers to be
discussed in the following. An additional factor $3$ due to their
three flavour over $\tau$ unique one may lead to respectively
($\leq 6\% $, $\leq 24\% $) of all HORTAUs events: a contribute
ratio that may be in principle an useful test to study the
balanced neutrino flavour mixing.

\subsection{Air Induced UHE $\nu$ Shower in EUSO}

UHE $\nu$ may hit an air nuclei and shower vertically or
horizontally or more rarely nearly up-ward: its trace maybe
observable by EUSO preferentially in inclined or horizontal
case.  Indeed  Vertical Down-ward  ($\theta \leq 60^o$) neutrino
induced Air Shower  occur mainly at lowest quota and they will
only partially shower their UHE $\nu$ energy because of the small
slant depth ($\leq 10^3 g cm^{-2}$) in most vertical down-ward
UHE $\nu$ shower. Therefore the observed  EUSO air mass ($1500
km^3$, corresponding to a $\sim 150$ $km^3$ for $10\%$ EUSO
record time) is only ideally the UHE neutrino calorimeter.
Indeed  inclined $\sim{\theta\geq 60^o }$) and horizontal
Air-Showers ($\sim{\theta\geq 83^o }$) (induced by GZK UHE
neutrino) may reach their maximum output  and their event maybe
observed ; therefore only a small fraction ($\sim 30\%$
corresponding to $\sim 50$ $km^3$ mass-water volume for EUSO
observation) of vertical downward UHE neutrino may be seen by
EUSO. This signal may be somehow hidden or masked by the more
common down-ward UHECR showers.  The key reading  signature will
be the shower height origination: $(\geq 40 km)$ for most
downward-horizontal UHECR,$(\leq 10 km)$ for most
inclined-horizontal Air UHE $\nu$ Induced Shower. A corresponding
smaller fraction ($\sim 7.45\%$) of totally Horizontal UHE
neutrino Air shower, orphan of their final Cherenkov flash, in
competition with the horizontal UHECR, may be also clearly
observed: their observable mass is only $V_{Air-\nu-Hor}$ $\sim
11.1$ $km^3$ for EUSO observation duty-cycle.  A more  rare, but
spectacular, double $\nu_{\tau}$-$\tau$ bang in Air (comparable
in principle to the PeVs expected  "double bang" in water
\cite{LearnedPakvasa1995}) may be exciting, but very difficult to
be observed.



The EUSO effective calorimeter mass for such Horizontal event is
only $10\%$ of the UHE $\nu$ Horizontal ones ($\sim 1.1$ $
km^3$); therefore its event rate is already almost excluded
needing a too high neutrino fluxes (see \cite{Fargion2002e});
indeed it should be also noted that the EUSO energy threshold
($\geq 3\cdot 10^{19}$eV) imply such a very large ${\tau}$
Lorents boost distance; such large ${\tau}$ track exceed (by more
than a factor three) the EUSO disk Area diameter ($\sim 450$km);
therefore the expected Double Bang Air-Horizontal-Induced ${\nu}$
Shower thresholds are suppressed by a corresponding factor. More
abundant single event Air-Induced ${\nu}$  Shower (Vertical or
Horizontal)  are facing different Air volumes and  quite
different visibility. It must be taken into account an additional
factor three (for the event rate) (because of three light
neutrino states) in the Air-Induced ${\nu}$  Shower arrival flux
respect to incoming $\nu_{\tau}$ (and $\bar{\nu_{\tau}}$ ) in
UPTAUs and HORTAUs, making the Air target not totally a
negligible calorimeter. The role of air Air-Shower will be
discussed elsewhere.


There are also a sub-category  of $\nu_{\tau}$ - $\tau$ "double
 bang" due to a first horizontal UHE $\nu_{\tau}$ charged current interaction
 in air  nuclei (the first bang) that is lost from the EUSO view;
 their UHE  secondary $\tau$ fly and decay leading to a Second Air-Induced Horizontal Shower, within the EUSO
 disk area.Therefore the total Air-Induced Horizontal Shower (for
 all $3$ flavours and the additional $\tau$ decay in flight) are summirized and considered
 in  Fig. \ref{fig:fig12} . The most relevant UHE neutrino signal, as discussed below, are due to the
 Horizontal Tau Air-Showers originated within the (much denser)Earth
 Crust: the called  HORTAUs (or Earth Skimming $\nu_{\tau}$).

\section{GZK $\nu_{\tau}-\tau$ Horizontal Showers in EUSO}

As already mention the UHE $\nu$ astronomy maybe greatly
amplified by $\nu_{\tau}$ appearance via flavour mixing and
oscillations. The consequent scattering of $\nu_{\tau}$ on the
Mountains or into the Earth Crust may lead to Horizontal Tau
Air-Showers :HORTAUs (or so called Earth Skimming Showers
\cite{Fargion2001a}~ \cite{Fargion2001b} ~\cite{Fargion2000-2002}
~\cite{Fengetall.2002}) . Indeed UHE $\nu_{\tau}$ may skip below
the Earth and escape as $\tau$ and finally decay in flight, within
air atmosphere, as well as inside  the Area of  view of EUSO, as
shown in Fig.\ref{fig:fig13} below.

\subsection{UPTAUs and HORTAUs effective Volumes}

\begin{figure}\centering\includegraphics[width=6cm]{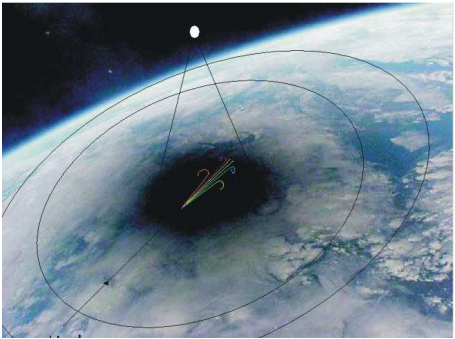}
\caption {A schematic Horizontal High Altitude Shower or similar
Horizontal Tau Air-Shower (HORTAUs) and its open fan-like jets
due to geo-magnetic bending seen from a high quota by EUSO
satellite. The image background is moon eclipse shadow observed
by Mir on Earth. The forked Shower is multi-finger containing a
inner $\gamma$ core and external fork spirals due to $e^+  e^-$
pairs (first opening) and  ${\mu}^+ {\mu}^-$ pairs.}
\label{fig:fig13}
\end{figure}

\begin{figure}\centering\includegraphics[width=6cm]{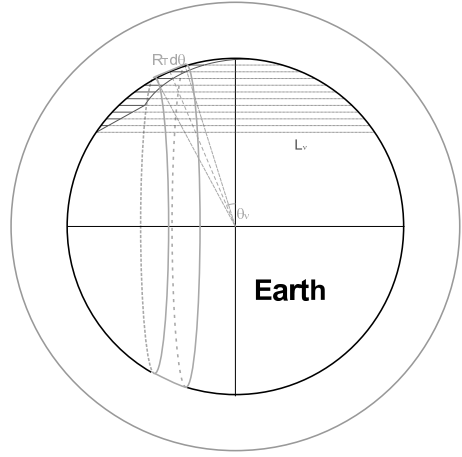}

\caption {A diagram describing the possible incoming neutrino
trajectory and the consequent outcoming tau whose further decay
in flight is the source of HORTAUs within the terrestrial
atmosphere.} \label{fig14}
\end{figure}

Any UHE-GZK Tau Air Shower induced event is approximately born
within a wide  ring  (whose radiuses extend between $R \geq 300$
and $R \leq 800$ km from the EUSO Area center). Because of the
wide area and deep $\tau$ penetration \cite{Fargion2000-2002}~
\cite{Fargion2002b} ~\cite{Fargion2002d} the amount of interacting
matter where UHE $\nu$ may lead to $\tau$ is huge ($\geq 2 \cdot
10^5$ $km^3$) ;however only a tiny fraction of these HORTAUs will
beam and Shower within the EUSO Area within EUSO. We estimate
(using also results in \cite{Fargion2000-2002}~
\cite{Fargion2002b} ~ \cite{Fargion2002c}~ \cite{Fargion2002d}
\cite{Fargion2002e} ) an effective Volumes for unitary area (see
schematic Fig. \ref{fig:fig15} below):



\begin{eqnarray}
\frac{V_{eff}}{A_{\oplus}}&=&\int_0^{\frac{\pi}{2}}\frac{(2\,\pi\,\,R_{\oplus}\cos\theta)
\,l_{\tau}\,\sin{\theta}}{4\,\pi\,R^2_{\oplus}}\cdot \nonumber\\
                          &\cdot&e^{-
\frac{2\,R_{\oplus}\,\sin{\theta}}{L_{\nu_{\tau}}}}\,R_{\oplus}\,d\theta\,=
\nonumber\\
&=&\frac{1}{2}\left({\frac{L_{\nu_{\tau}}}{2\,R_{\oplus}}}\right)
^2\,l_{\tau}\int_0^{\frac{2\,R_{\oplus}}{L_{\nu_{\tau}}}}t\cdot
e^{-\,t}d\,t
\end{eqnarray}

Where $V_{eff}$ is the effective volume where Ultra High Energy
neutrino interact while hitting the Earth and lead to escaping
UHE Tau: this volume encompass a wide crown belt, due to the
cross-section of neutrino Earth skimming along a ring of variable
radius, $R_{\oplus}\cos\theta$, and a corresponding skin crown of
variable depth $l_{\tau}$. $A_{\oplus}$ is the total terrestrial
area, $l_{\tau}$ is the tau interaction lenght, $L_{\nu_{\tau}}$
is the Ultra High Energy Neutrino tau interaction (charged
current) in matter . The resulting detectable volume from EUSO
becomes,(see the combined effective volume in  Fig.
\ref{fig:fig15}):

\begin{eqnarray}
V_{eff}&=&\frac{1}{2}\,A_{Euso}\,\left({1-e^{-
\frac{L_0}{c\,\tau_{\tau}\,\gamma_{\tau}}}}\right)
\,\left({\frac{L_{\nu_{\tau}}}{2\,R_{\oplus}}}\right)^2\cdot
\nonumber\\
          &\cdot&l_{\tau}\left[{1\,-\,e^{-
\frac{2\,R_{\oplus}}{L_{\nu_{\tau}}}}(1\,+\,\frac{2\,R_{\oplus}}{L_{\nu_{\tau}}})
}\right]
\end{eqnarray}


The above effective volume should be considered for any given
neutrino flux to estimate the outcoming EUSO event number
\cite{Fargion2003}. These general expression will be plot assuming
a minima GZK neutrino flux $ \phi_{\nu}$ just comparable to
observed UHECR one $ \phi_{\nu}\simeq \phi_{UHECR} \simeq 3\cdot
10^{-18} cm^{-2} s^{-1} sr{-1}$ at the same energy ($E_{\nu}=
E_{UHECR}\simeq 10^{19} eV$). This assumption may changed at will
(model dependent) but the event number will scale linearly
accordingly to any incoming neutrino flux model. From here we may
estimate the event rate in EUSO by a simple extension:

$N_{eventi}\,=\,\Phi_{\nu}\,4\,\pi\,\eta_{Euso}\Delta
\,t\,\left({\frac{V_{eff}}{L_{\nu}}}\right)$.

Where $\eta_{Euso}$ is the  duty cycle fraction of EUSO,
$\eta_{Euso} \simeq 10\%$, $\Delta \,t\ \simeq 3$  $years$ and
$L_{\nu}$ has been defined in Fig.\ref{fig:fig8}.


\begin{eqnarray}
N_{eventi}&=&\Phi_{\nu_\tau}\,A_{Euso}\,(\frac{1}{2}\,4\pi\,\eta_{Euso})\,{\Delta{t}}\cdot \nonumber\\
          &\cdot&\left({1-e^{-\frac{L_0}{c\,\tau_{\tau}\,\gamma_{\tau}}}}\right)\left({\frac{l_{\tau}}{L_{\nu_{\tau}}}}\right)\left({\frac{L_{\nu_{\tau}}}{2\,R_{\oplus}}}\right)^2\cdot \nonumber\\
          &\cdot&\left[{1\,-\,e^{-\frac{2\,R_{\oplus}}{L_{\nu_{\tau}}}}(1\,+\,\frac{2\,R_{\oplus}}{L_{\nu_{\tau}}})}\right]
\end{eqnarray}

The term
$\left({1-e^{-\frac{L_0}{c\,\tau_{\tau}\,\gamma_{\tau}}}}\right)$
takes into account the limit of the air atmosphere on Earth which
is basically at height $h\simeq 23 km$ and at a corresponding
distance $L_0 \simeq 600 km$. It should be remind that all these
event number curves for EUSO are already suppressed by a factor
$\eta_{Euso} \simeq 0.1$ due to minimal EUSO duty cycle.

The same effective Volume , under the assumptions of an incoming
neutrino energy $E_{\nu_{\tau}}=4 E_{\tau}$ and under the
assumption that the outcoming Tau energy leading to a Shower is
$E_{\tau}= 1.5 E_{Shower}$  corresponds, in water at $E_{\tau}=
10^{19} eV$ , to $V_{eff}=1.13 \cdot 10^{3} km^3$; the
corresponding event number is $N_{ev} = 13.4$. See
Fig.\ref{fig:fig15}.

As it is manifest from the above curve the maximal event numbers
takes place at EeV energies. Therefore from here we derived the
primary interest for EUSO to seek lowest threshold.

\begin{figure}\centering\includegraphics[width=4cm, angle=270]{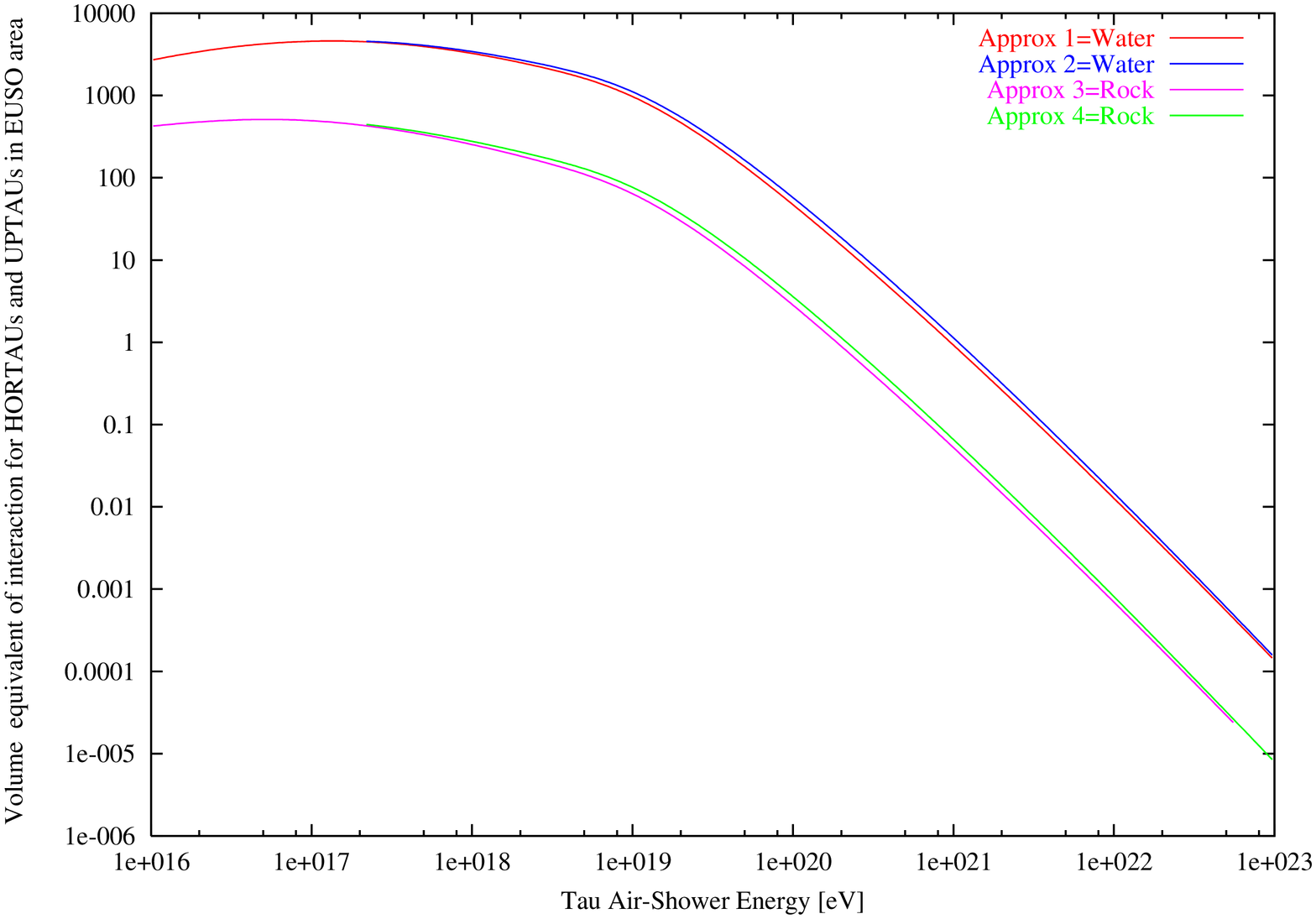}
\caption {Effective Volume of EUSO Event for UPTAUs and HORTAUs
assuming a GZK neutrino flux  $\frac{dN_{\nu}}{dE_{\nu} dt dA
d\Omega}\cdot E_{\nu} = 3\cdot 10^{-18} cm^{-2} s^{-1} sr^{-1}$,
corresponding to fluence $\Phi_{\nu} = 30 eV cm^{-2} s^{-1}
sr^{-1}$. The effective Volume is  for all Upcoming
Upward-Horizontal Tau Air-Showers, contained within the EUSO area
nearly at $25$ km altitude where it may still shower. This Volume
is estimated under the assumptions of an incoming neutrino energy
$E_{\nu_{\tau}}= 4 E_{\tau}$ and under the assumption that the
outcoming Tau energy is  leading to a Shower is $E_{\tau}= 2
E_{Shower}$ } \label{fig:fig15}

\centering\includegraphics[width=4cm,
angle=270]{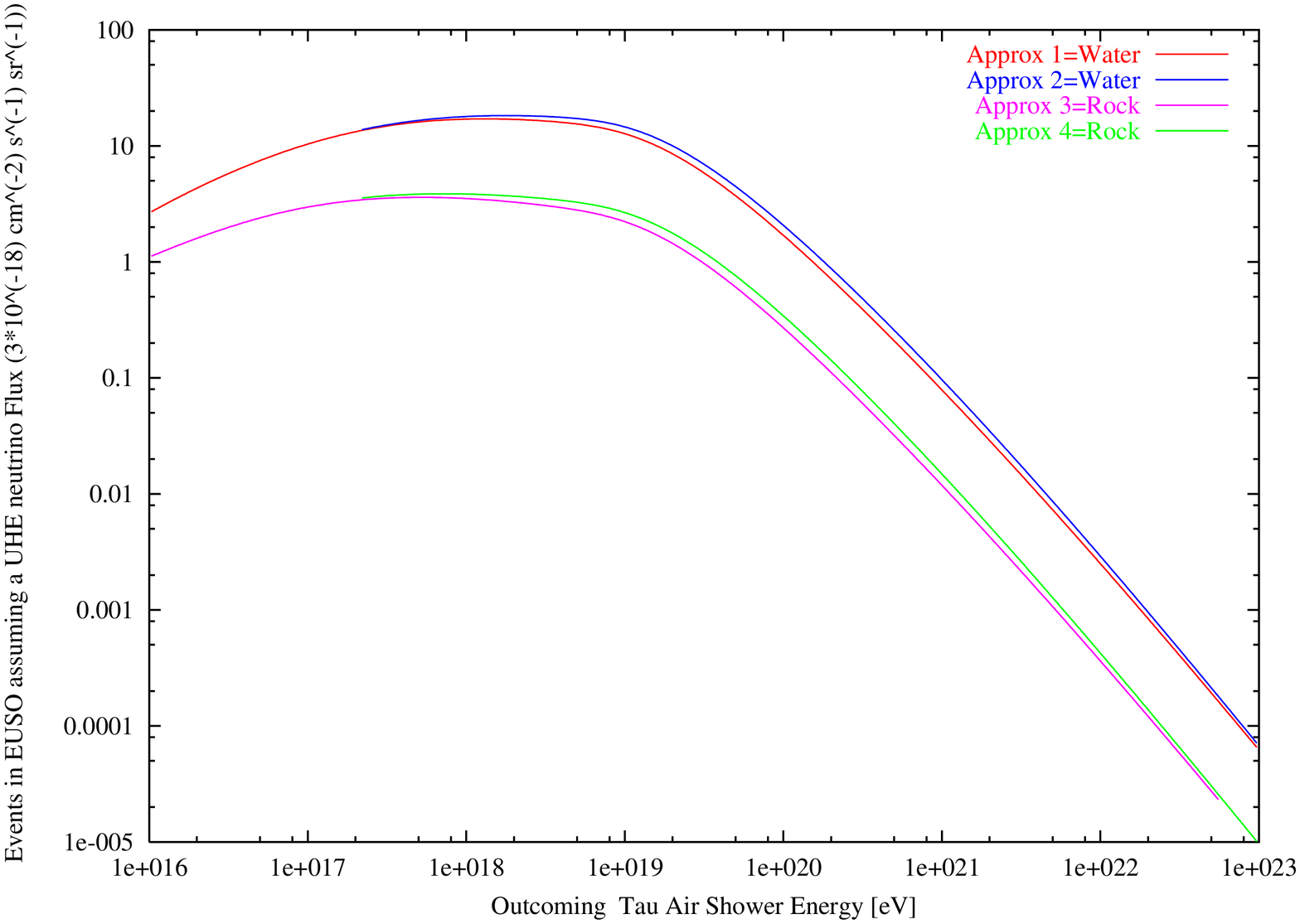}
\caption {Number of EUSO Event for HORTAUs in 3 years record
assuming a GZK neutrino flux $\frac{dN_{\nu}}{dE_{\nu} dt dA
d\Omega}\cdot E_{\nu} = 3\cdot 10^{-18} cm^{-2} s^{-1} sr^{-1}$.
The incoming neutrino has an energy $4$ larger the outcoming Tau;
this born Tau has an energy $2$ times the final Tau Air-Shower
end energy.} \label{fig:fig16}
\end{figure}
\vspace{0.3cm}
\begin{figure}
\centering\includegraphics[width=4cm,angle=270]{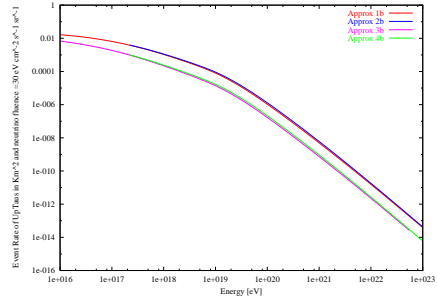}
\caption {The Number of  Events for HORTAUs and UPTAUs in 3 years
record for square km$^2$ area assuming a given flat GZK neutrino
fluence $\Phi_{\nu} = 30 eV cm^{-2} s^{-1} sr^{-1}$ in all
energies. This value corresponds to a GZK minimal flux and a
minimal UHE neutrino fluence comparable to  the Waxman-Bachall
limit. The assumed incoming neutrino has an energy $4$ larger
than the outcoming Tau; this born Tau has an energy $2$ times the
final Tau Air-Shower end energy. Note however that at lowest
energies the present approximation should slightly deviate
because the total exact Earth opacity  has been a little
underestimated but the $\nu_{\tau}- \tau$ regeneration has been
totally neglected. Nevertheless this estimate in present form is a
realistic approximation in the central  energy window considered.}
\label{fig:fig20}
\end{figure}
\vspace{0.3cm}

\subsection{UPTAUs-HORTAUs unified volume }

The above expression for the horizontal tau air-shower contains ,
at lowest energies, the UPTAUs case. Indeed it is possible to see
that the same above  effective volume at lowest energies simplify
and  reduces to:

$V_{eff}=\frac{1}{2}\,A_{Euso}\,\left({1-e^{-\frac{L_0}{c\,\tau_{\tau}\,\gamma_{\tau}}}}\right)\,l_{\tau}$
Because one observes the Earth only from one side   the Area
factor in eq. $1$ should be $A_{\oplus} = {2\,\pi\,R^2_{\oplus}}$
and therefore the half in above formula may be dropped and the
final result is just the common expression $V_{eff} =
A_{Euso}\,l_{\tau}$.







\begin{figure}
\centering\includegraphics[width=8cm]{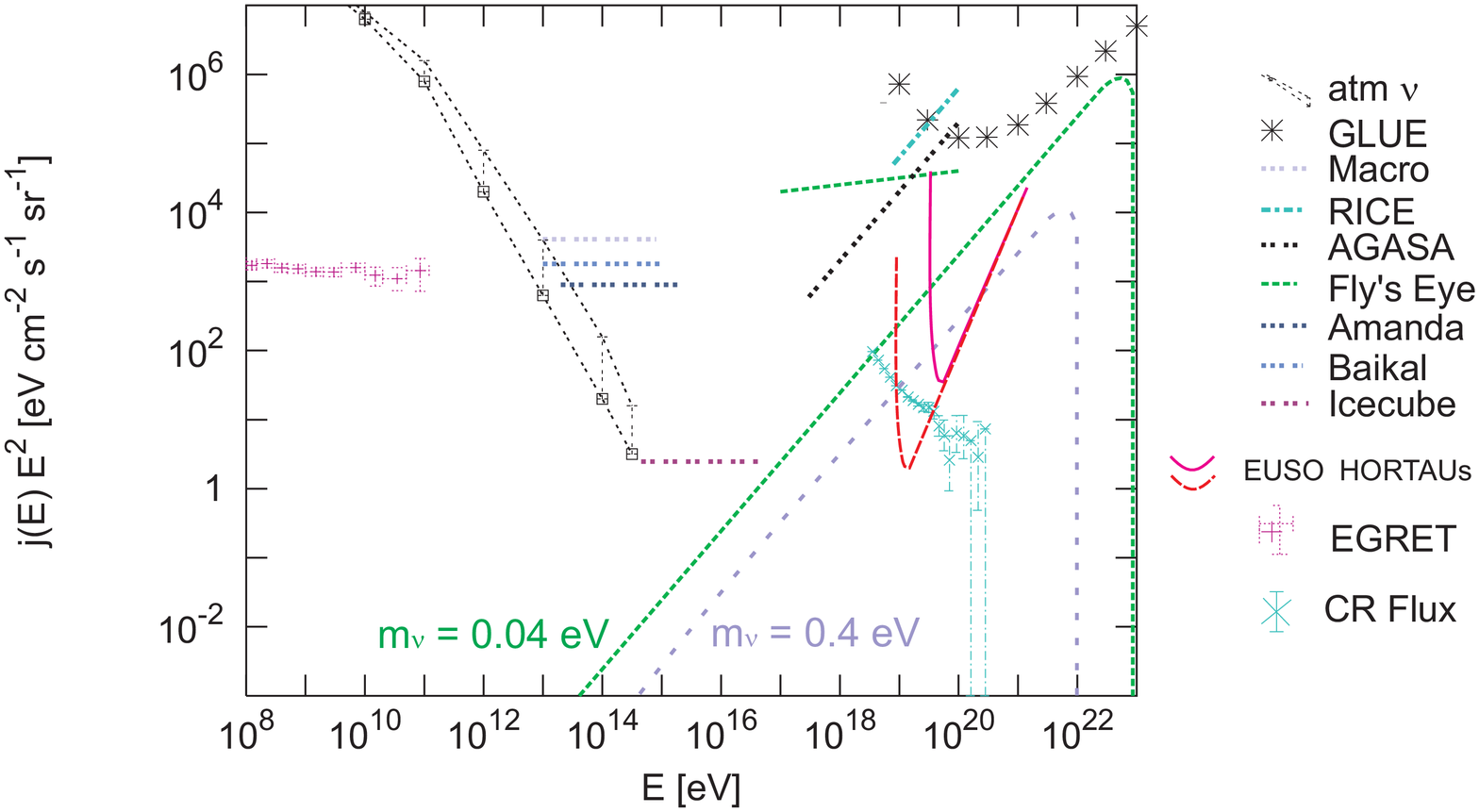}
\vspace{-1.1cm}
\caption {EUSO thresholds for Horizontal Tau Air-Shower HORTAUs
(or Earth Skimming Showers) over few $\gamma$, $\nu$ and Cosmic
Rays (C.R.) Fluence and bounds. Continuous and dashed curve for
HORTAUs are drawn assuming respectively an EUSO threshold at
$3\cdot 10^{19}$eV and at lower $10^{19}$eV values. Because the
bounded $\tau$ flight distance (due to the contained terrestrial
atmosphere height) the main signal is better observable at $1.1
\cdot 10^{19}$eV than higher energies. The Fluence threshold for
EUSO has been estimated for a three year experiment lifetime.
Z-Shower or Z-Burst expected spectra in light neutrino mass values
($m_{\nu} = 0.04, 0.4$ eV) are shown. } \label{fig:fig17}

\vspace{0.3cm}
\centering
\includegraphics[width=8cm]{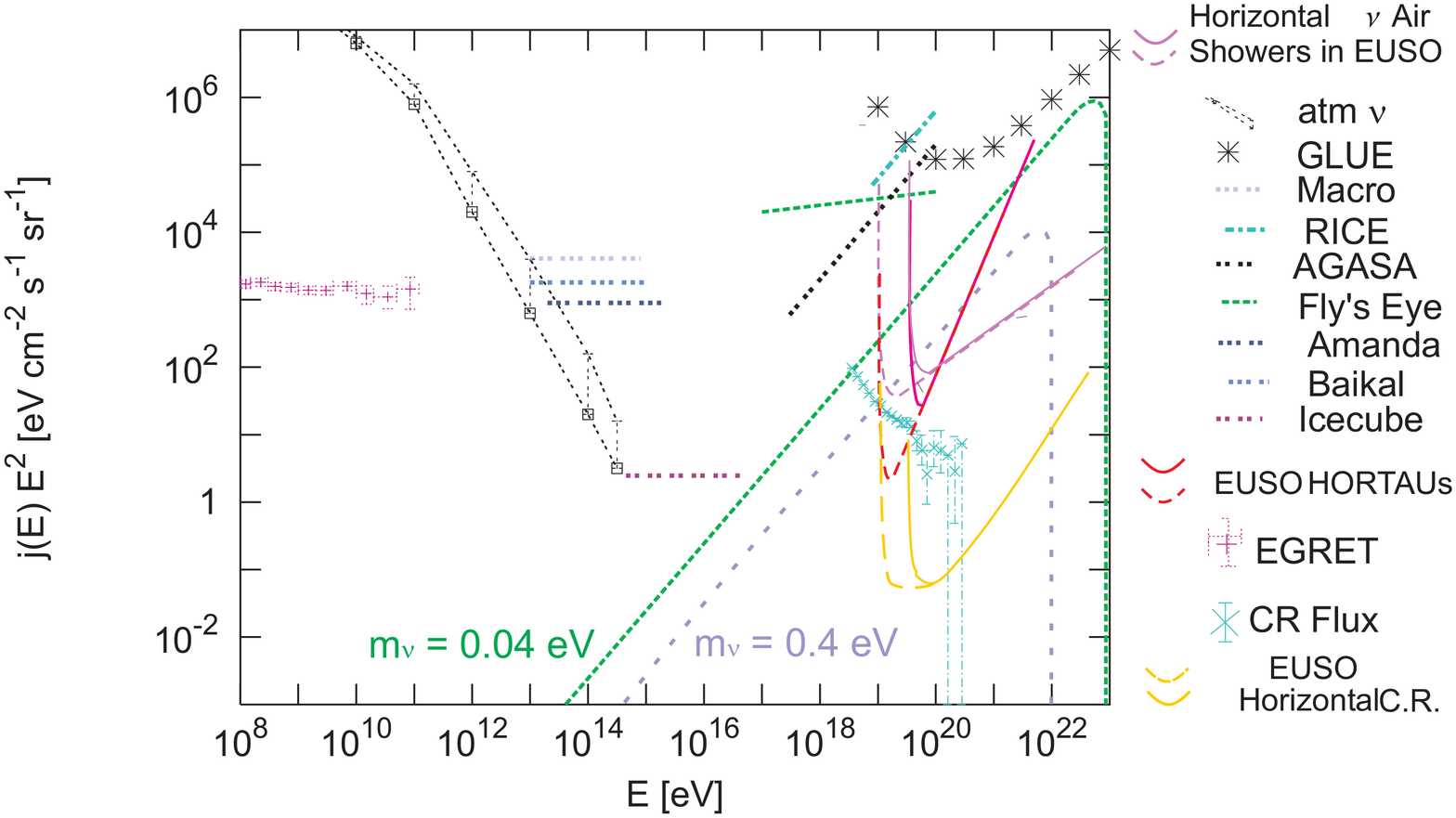}
 \vspace{-1cm}
\caption {EUSO thresholds for Horizontal Tau Air-Shower shower,
HORTAUs (or Earth Skimming Showers) over all other $\gamma$, $\nu$
and Cosmic Rays (C.R.) Fluence and bounds. The Fluence threshold
for EUSO has been estimated for a three year experiment lifetime.
Competitive experiment are also shown as well as the Z-Shower
expected spectra in light neutrino mass values ($m_{\nu} = 0.04,
0.4$ eV). As above dash curves for both HORTAUs and Horizontal
Cosmic Rays are drawn assuming an EUSO threshold at $10^{19}$eV.
} \label{fig:fig18}
\end{figure}





\subsection{Partially, Fully contained  events in EUSO}
  The HORTAUs are very long showers. Their lenght may exceed
  two hundred kilometers. This trace may be larger than the EUSO
  radius of Field of View. Therefore there may be both contained
  and partially contained events. There may be also crossing
  HORTAUs at the edges of EUSO  disk area. However most
  of the events will be partially contained, either just on their
  birth or at their end, equally balanced in number. Because of the HORTAU Jet forked shower,
  its up-going direction, its fan like structure, these partially
  contained shower will be the manifest and mostly useful and
  clear event. The area of their origination, four times larger than
  EUSO field of view, will be mostly
  outside the same EUSO area. Their total number count will double the
  event rate $N_{ev}$ (and the corresponding $V_{eff}$) of
  HORTAUs. The additional crossing event will make additional events (a small fraction) of the effective
  volume of HORTAUs at $10^{19} eV$ the most rich neutrino signal few times larger the Air induced events.
  The same doubling will apply only to UHECR horizontal shower
  while the downward Ultra High Energy Neutrino will not share
  this phenomena (out of those $\simeq 6\%$ of the Horizontal Air Neutrino
  Shower).

 \section{Conclusions}
The Neutrino Astronomy may be widely discovered by Upward and
Horizontal $\tau$ Air-Showers. The Tau neutrinos , born
abundantly by flavour mixing will probe such Astronomy above PeVs
up to EeVs energies, where astrophysics rule over atmospheric
neutrino noise. The same UHE $\overline{\nu_{e}}$  at $E_{\nu_{e}}
= \frac{{M_W}^2}{2 \cdot m_e} \simeq 6.3 PeV $  must be a peculiar
neutrino astronomy born beyond Mountain Chains
\cite{Fargionetall.1999}~\cite{Fargion2000-2002} with its very
distinctive signature. This ground $\tau$ Air-Showers astronomy
may test (by shower distance correlation) very deeply the
$\overline{\nu_{e}}$ versus $\overline{\nu_{\tau}}$,
${\nu_{\tau}}$ flavour fluxes. Past detectors as GRO BATSE
experiment might already found some direct signals of such rare
UPTAUs or HORTAUs; indeed their observed Terrestrial Gamma Flash
event rate translated into a neutrino induced  upward shower (see
Fig.\ref{fig:fig19}) leads to a most probable flux both at PeVs
energies  just at a level comparable to most recent AMANDA
threshold sensitivity: for horizontal TGF events at $10^{19}$ eV
windows, these signals just fit the Z-Burst model needed fluence
(for neutrino  at $0.04-0.4$ eV masses).
 Future EUSO telescope detector, if little enlarged will easily probe even the smallest, but necessary
Neutrino GZK fluxes with clear sensitivity (see Fig.
\ref{fig:fig17},\ref{fig:fig18}). We therefore expect that a
serial of experiment will soon turn toward this last and
neglected, but most promising Highest Energy Neutrino Tau
Astronomy searching for GZK or Z-Showers neutrino signatures just
beyond the horizon.

\subsection{Acknowledgment}
The author thanks for useful numerical and editorial support Pier
Giorgio De Sanctis Lucentin , Cristina Leto and Massimo De Santis.

\bibliography{xbib}
\end{document}